\documentclass[11pt,a4paper,final]{article}
\usepackage[left=1in,right=1in,top=1in,bottom=1in]{geometry}
\usepackage{authblk}
\PassOptionsToPackage{sort&compress}{natbib}
\usepackage[numbers]{natbib}
\usepackage{booktabs} 
\usepackage[ruled]{algorithm2e} 
\usepackage{tikz}
\usetikzlibrary{arrows}
\usetikzlibrary{shapes.multipart}
\usepackage{caption}

\SetAlFnt{\small}
\SetAlCapFnt{\small}
\SetAlCapNameFnt{\small}
\SetAlCapHSkip{0pt}
\IncMargin{-\parindent}
\usepackage[all]{nowidow}

\usepackage[font=small,labelfont=bf]{caption}
\usepackage[utf8]{inputenc} 
\usepackage[T1]{fontenc}    
\usepackage{xcolor}
\usepackage{color}
\usepackage[pagebackref=true,breaklinks=true,colorlinks,bookmarks=false]{hyperref}
\definecolor{deepred}{HTML}{940000}
\hypersetup{linkcolor=deepred}
\hypersetup{citecolor=[rgb]{0.4,0.15,0.95}}
\usepackage{url}            
\usepackage{booktabs}       
\usepackage{amsfonts}       
\usepackage{nicefrac}       
\usepackage{microtype}      
\usepackage{xcolor}         
\usepackage{amsmath}
\usepackage{mathtools}
\usepackage{amsthm}
\usepackage{amsmath}
\usepackage{amssymb}
\usepackage{bbm}
\usepackage{tikz} 
\usepackage{wrapfig}
\usepackage{floatrow}
\usepackage{enumitem}

\usepackage{caption}
\usepackage{subcaption}

\usepackage{color-edits}
\addauthor{hs}{blue}
\addauthor{ub}{red}

\newcommand*\samethanks[1][\value{footnote}]{\footnotemark[#1]}

\title{When Should Algorithms Resign? A Proposal for AI Governance}

\author[1,2]{Umang Bhatt\thanks{Both authors contributed equally. Correspondence to: \href{mailto:umangbhatt@nyu.edu}{umangbhatt@nyu.edu}} }
\author[3,4]{Holli Sargeant\samethanks}
\affil[1]{Center for Data Science, New York University}
\affil[2]{The Alan Turing Institute}
\affil[3]{Faculty of Law, University of Cambridge}
\affil[4]{Berkman Klein Center for Internet \& Society}
\date{}

\begin{document}

\maketitle

\begin{abstract}
Algorithmic resignation is a strategic approach for managing the use of artificial intelligence (AI) by embedding governance directly into AI systems. It involves deliberate and informed disengagement from AI, such as restricting access AI outputs or displaying performance disclaimers, in specific scenarios to aid the appropriate and effective use of AI. By integrating algorithmic resignation as a governance mechanism, organizations can better control when and how AI is used, balancing the benefits of automation with the need for human oversight.
\end{abstract}

\section*{Introduction}
Technological progress often leads to periods of rapid adoption before the full effects on users are well understood.
To mitigate the potential abuse of technology, policies, norms, and laws may be implemented to promote responsible use and non-use.
For example, speed limits in the UK were introduced three years after cars were deemed fast enough to be dangerous~\citep{wilmot1999speed}; the first Food and Drug Administration court ruling came 46 years after it began analyzing samples of food and soil~\citep{basile2004FDA}; and video game ratings came 18 years after parents became concerned with the increasing levels of violence in games and their effects on children~\citep{stroud2008videogames}.
\textbf{Precise discussions around the use and non-use of emerging technologies are nonnegotiable to prevent misuse and abuse.}

Tesla recently won a pivotal lawsuit related to its AI-powered Autopilot system. 
Jurors held that a fatal accident was not the fault of Tesla's driver-assistance system because all drivers were informed they must maintain full control over the vehicle~\citep{TeslaCase}. 
The driver was ultimately held responsible, and the verdict absolves Tesla's Autopilot system at a pivotal moment for autonomous vehicle technology. 
Importantly, this case raises critical questions about the complex interplay between human operators and AI systems and underscores the legal challenges of humans and AI systems working together. 

Organizations that grant their members access to AI systems are responsible for its use. 
While some members may use AI judiciously, human oversight does not inherently guarantee proper use. 
Even partial automation may lull some members into complacency. 
Members who become complacent expose organizations to unforeseen risks, including financial loss or reputational damage.
Research shows that over-reliance on AI systems often leads to worse performance on tasks compared to either the user or the AI system working alone~\citep{zhang2020}.

Unlike traditional ex-post enforcement of non-use, we advocate building governance mechanisms into AI systems directly. 
We propose \textbf{algorithmic resignation}, where organizations deliberately limit algorithmic assistance in favor of (unaided) human decision-making. 
This approach embeds governance into system design, providing concrete guidance for appropriate use and distinguishing permissible levels of automation from inappropriate or unlawful ones.
Algorithmic resignation allows organizations to control when and how AI systems are used.   

\section*{What is Algorithmic Resignation?}
Algorithmic resignation represents a strategic shift in how organizations approach responsible AI by selectively disengaging AI systems in particular scenarios to enhance decision-making. 
Unlike blanket strategies that either fully integrate or entirely reject AI use, algorithmic resignation embeds governance mechanisms to ensure appropriate and effective use.
Organizations can incorporate algorithmic resignation directly into the AI systems they build, procure, or deploy. 
We urge organizations to proactively decide when and where AI should be curtailed or allowed, based on specific, predefined criteria, considering both operational needs and legal considerations. 

Algorithmic resignation is distinct from other research on the social phenomenon where individuals feel helpless to the pervasive monitoring and data collection practices of digital entities.
Instead, algorithmic resignation turns the focus onto active and strategic choices to limit the role of AI systems in certain contexts. 
Controlled use of AI can mitigate risks and enhance human oversight, providing a necessary counterbalance in environments where AI's decision-making capabilities are powerful but not infallible. 
Deciding when to mandate the non-use of AI systems, whether temporary or permanent, depends on various factors, including the system's performance, the user's role and expertise, and the sociotechnical and legal contexts.

\paragraph{System Performance.} One set of factors for resignation may depend on the properties of the AI system itself. 
When systems encounter scenarios outside their training data or when systems are highly uncertain in their predictions, resignation can act as a safeguard against displaying potentially erroneous predictions. 
Existing tools in the machine learning community, such as deferral mechanisms that abstain from providing predictions on specific data, can be repurposed to implement algorithmic resignation~\cite{bartlett2008classification}. 
Well-calibrated AI systems that are confident when correct and uncertain when incorrect can instantiate resignation by ignoring uncertain predictions in favor of human judgment.
Displaying uncertain AI system predictions to members would likely degrade member performance on various decision-making tasks.
Organizations will need to decide on a threshold beyond which AI systems should not be accessible, providing a \textit{veil of selectivity} that empowers members to make decisions in specific settings without assistance from potentially misleading AI systems.

\paragraph{User Expertise.} Beyond factors that depend on the AI system itself, organizations can consider resignation based on the user's expertise.
As preferences and expertise may differ across members, organizations can enable each member to access AI systems when most appropriate for them.
Junior doctors, for instance, may benefit from assistance more than senior physicians; however, even experienced doctors might be nudged to view AI predictions to counteract overconfidence~\citep{sunstein2023decisions}. 
Resignation protects organizations from members who overestimate their expertise by forcing AI system assistance upon them or those who undermine their abilities by revoking AI system access. 
AI systems can also resign in ways that complement, not reinforce, a member's skills.
One practical application could be selectively enabling non-native English speakers to use large language models, like ChatGPT, for composing clearer communications, while being prohibited from using them for critical decision-making tasks. 

\paragraph{Sociotechnical Information.} Beyond individual personalization, organizations can establish broad policies for algorithmic resignation based on internal norms (e.g., documentation practices) or external industry-wide regulations (e.g., professional conduct rules).
A law firm might allow the use of ChatGPT for internal research or communications but not for external client interactions, or for routine tasks by paralegals but not for attorneys' legal advisory work.
The use of an AI system should not interfere with lawyers' responsibility to deliver competent legal services, although using new technology may be relevant to upholding their duty of competence.
Organizations can craft a strategy to decide how much AI assistance to provide to each member.

\begin{figure*}
    \centering
    \includegraphics[width=\textwidth]{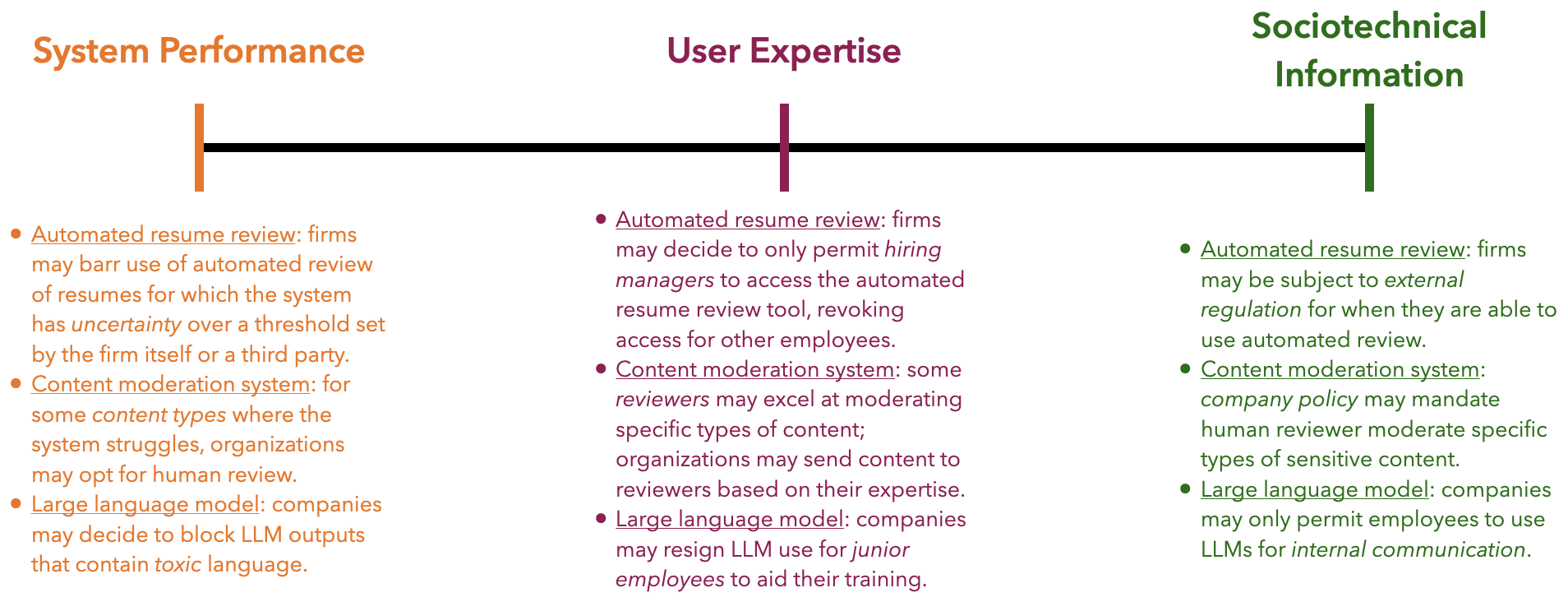}
    \caption{Examples of factors influencing algorithmic resignation across an organization.}
    \label{fig:levelofaccess}
\end{figure*}

Operationally, algorithmic resignation can be implemented in various ways, from completely barring access to AI outputs in certain scenarios to softer measures like explicit disclaimers or guidance. 
For instance, a large language model might refuse to provide responses to queries that border on legal advice, instead offering unregulated general legal information. 
Figure \ref{fig:levelofaccess} provides examples of factors influencing the use of algorithmic resignation.  
By promoting the non-use of AI systems through resignation, organizations can prevent misuse or, worse, abuse of AI-assisted decision-making.

\section*{Cases for Algorithmic Resignation}

Algorithmic resignation offers a range of benefits spanning economic, reputational, and legal domains. 
These advantages not only improve operational efficiency but also align with broader ethical and regulatory expectations.

\paragraph*{Financial Benefit.} 
Algorithmic resignation can lead to significant cost savings and increased efficiency. 
By selectively using AI, organizations can optimize decision-making processes, ensuring AI systems are only employed where they add genuine value and avoiding costly errors associated with over-reliance on AI.
Providing structured choices on when to use AI helps members exercise better control and judgment, aligning their actions with the organization's best interests~\citep{jensen_theory_1976}.
Selectivity optimizes overall performance, minimizing friction and costs arising from divergent interests among stakeholders. 
Beyond improving decision-making accuracy and cost-effectiveness, organizations can enhance members' understanding of when AI assistance may hinder their performance and improve third parties' perceptions of AI usage within the organization.
    
\paragraph*{Reputational Gain.} Implementing algorithmic resignation demonstrates a commitment to responsible AI use, setting a precedent for meaningful commitment to the conscious application of technology. 
Such commitment enhances trustworthiness, as stakeholders and the broader public observe the organization acting in \textit{good faith}. 
By governing AI use through algorithmic resignation, stakeholders are assured that AI systems supplement rather than substitute human judgment.
The reputational risks associated with AI misuse are significant and well-documented, often leading to legal scrutiny and damage to the organization's image. 
A proactive stance through resignation mitigates these risks and contributes to building a sustainable and positive brand image over in the long term.

\paragraph*{Legal Compliance.} In the face of increasing calls for external AI regulation, organizations have a strong incentive to develop internal governance mechanisms like algorithmic resignation. 
Algorithmic resignation allows organizations to self-regulate meaningfully by embedding governance mechanisms directly within AI systems.
By doing so, organizations can proactively address emerging legal requirements, aligning with and anticipating external regulatory demands such as the European Union's Artificial Intelligence Act and the US Executive Order on AI~\citep{EUAIAct,EO14110}. 
Algorithmic resignation not only aids in compliance but also reduces the risk of legal challenges stemming from over-reliance on AI, such as errors, privacy breaches, or unethical decision-making. 
As legislation evolves, adopting internal mechanisms to ensure appropriate and effective AI use becomes critical for organizations.

\section*{Considerations for Implementing Algorithmic Resignation}

The successful deployment of algorithmic resignation requires careful consideration of several factors to ensure alignment with organizational goals and regulatory requirements.

\paragraph*{Incentive Misalignment.} Understanding the various incentives driving each stakeholder is crucial for the successful implementation of algorithmic resignation. 
Managers or employees may have personal motivations that differ from those of the organization's directors or owners, leading to blatant or careless misuse of AI systems. 
Economic costs can arise from inefficiencies or losses due to misaligned objectives. 
Differing incentives mean that each stakeholder may desire a different level of AI use. 
Effective management requires oversight and incentives to encourage behavior that aligns with the organization's best interests. 
For instance, in complex healthcare settings, the decision to use tools like  electrocardiograms  can be influenced by various stakeholders' interests. 
Hospitals might favor the use of AI for efficiency and improved diagnostics, while insurers might be cautious about overuse due to high costs. 
Balancing these incentives is a key challenge in implementing AI systems effectively.

\paragraph*{Nudging over Resigning.} In some contexts, resigning the use of AI systems for certain members may be too strong or lead to issues of disparate treatment. 
Organizations might opt instead to discourage AI use in specific scenarios through nudging. 
Behavioral economics supports the idea of dissuasion rather than prohibition, gently steering behavior away from less favorable outcomes~\citep{thaler2009nudge}. 
Nudging can be implemented through restraints, disclosures, and guidance to members using AI systems. 
For example, social media companies might embed a warning in an automated content moderation tool indicating high uncertainty and suggesting manual review. 
Nudging, like resigning, prevents over-reliance on AI while still capitalizing on its benefits.

\paragraph*{Level of Engagement.} The extent to which members use AI systems in their decision-making processes varies, influencing the implementation of algorithmic resignation. 
In settings where AI significantly influences outcomes, high engagement may conflict with regulatory oversight. 
For instance, organizations operating in the EU must ensure members are not making solely automated decisions with significant effects on individuals unless within specific General Data Protection Regulation (GDPR) conditions~\citep{GDPR}. 
Even if members participate in the decision-making process, it is considered solely automated if they cannot influence the causal link between the automated processing and the final decision.
For decisions subject to regulation, resignation provides restraint by barring access to AI to ensure the appropriate level of engagement. 
Where AI engagement is high, there may be regulatory needs for thresholds between automation and human oversight, and resignation can help achieve this balance.

\section*{Conclusion}

The implementation of algorithmic resignation requires a nuanced understanding of its potential benefits, stakeholder incentives, and the level of user engagement with AI. 
By carefully considering these aspects, organizations can develop strategies that ensure the appropriate and effective use of AI. 
Algorithmic resignation offers a balanced approach, aligning AI use with broader organizational objectives and ethical standards. 
This proactive stance not only mitigates risks associated with AI misuse but also enhances trust, compliance, and operational efficiency. 
By embedding governance directly into AI systems, organizations can foster a culture of responsible AI use, paving the way for sustainable and ethical technological advancement.

\subsection*{Acknowledgements}
The authors thank Katherine Collins, Adrian Weller, John Zerilli, and anonymous reviewers for their helpful feedback. 
UB acknowledges funding from the Alan Turing Institute.
HS acknowledges funding from the General Sir John Monash Foundation.




\bibliography{papers}
\bibliographystyle{IEEEtran}

\end{document}